\documentclass[aip,apl,preprint]{revtex4-2}

\usepackage{amsmath}
\usepackage{graphicx}
\usepackage{dcolumn}	
\usepackage{hyperref}
\usepackage[T1]{fontenc}
\usepackage{textcomp}
\usepackage{bm}				
\usepackage{xfrac}

\begin{document}

\title{Magnetoelectric Coupling in Pb(Zr,Ti)O$_3$/CoFeB Nanoscale Waveguides Studied by Propagating Spin-Wave Spectroscopy}
\author{Daniele Narducci}
\affiliation{Imec, 3001 Leuven, Belgium}
\affiliation{KU Leuven, Departement Materiaalkunde (MTM), 3001 Leuven, Belgium}
\author{Xiangyu Wu}
\affiliation{Imec, 3001 Leuven, Belgium}
\author{Isabella Boventer}
\affiliation{Unit\'e  Mixte de Physique CNRS, Thales, Universit\'e Paris-Saclay, Palaiseau 91767, France}
\author{Jo De Boeck}
\affiliation{Imec, 3001 Leuven, Belgium}
\affiliation{KU Leuven, Departement Elektrotechniek (ESAT), 3001 Leuven, Belgium}
\author{Abdelmadjid Anane}
\affiliation{Unit\'e  Mixte de Physique CNRS, Thales, Universit\'e Paris-Saclay, Palaiseau 91767, France}
\author{Paolo Bortolotti}
\affiliation{Unit\'e  Mixte de Physique CNRS, Thales, Universit\'e Paris-Saclay, Palaiseau 91767, France}
\author{Christoph Adelmann}
\affiliation{Imec, 3001 Leuven, Belgium}
\author{Florin Ciubotaru}
\affiliation{Imec, 3001 Leuven, Belgium}

\begin{abstract}
This study introduces a method for the characterization of the magnetoelectric coupling in nanoscale Pb(Zr,Ti)O$_3$/CoFeB thin film composites based on propagating spin-wave spectroscopy. Finite element simulations of the strain distribution in the devices indicated that the magnetoelastic effective field in the CoFeB waveguides was maximized in the Damon--Eshbach configuration. All-electrical broadband propagating spin-wave transmission measurements were conducted on Pb(Zr,Ti)O$_3$/CoFeB magnetoelectric waveguides with lateral dimensions down to 700 nm. The results demonstrated that the spin-wave resonance frequency can be effectively modulated by applying a bias voltage to Pb(Zr,Ti)O$_3$. The modulation is hysteretic due to the ferroelastic behavior of Pb(Zr,Ti)O$_3$. An analytical model was then used to correlate the change in resonance frequency to the induced magnetoelastic field in the magnetostrictive CoFeB waveguide. We observe a hysteresis magnetoelastic field strength with values as large as $5.61$ mT, and a non-linear magnetoelectric coupling coefficient with a maximum value of $1.69$ mT/V.
\end{abstract}

\maketitle

Magnetoelectric materials possess the remarkable ability to control their magnetic state through the application of electric fields. Among magnetoelectric systems, single-phase multiferroic materials show an intrinsic coupling between (anti-)ferromagnetic and ferroelectric properties.\cite{Spaldin2005} Multiferroic materials have recenrly received much attention due to their potential for ultralow-energy manipulation of the magnetization by voltages,\cite{Lin2019} with potential applications in transducers in emerging spintronic devices, \emph{e.g.}, in MESO logic,\cite{Manipatruni2019} spin-wave devices, \cite{Cherepov2014} or as switching mechanism in magnetoelectric MRAM devices.\cite{Bibes2008} 

However the implementation of multiferroics in practical devices is limited for many materials by low Curie temperatures, limiting their use to cryogenic temperatures,\cite{Hill2000} and by the generally rather weak magnetoelectric coupling.\cite{Khomskii2009} For this reason, magnetoelastic compounds have emerged as alternatives.\cite{Vopson2017} These materials are composed of piezoelectric  and magnetostrictive (or piezomagnetic) layers. Applying an electric field to the piezoelectric layer induces mechanical stress, which is transferred to the magnetostrictive layer and exerts a torque on its magnetization. Hence, in these compounds, the magnetoelectric coupling is mediated by strain. This has been, \emph{e.g.}, employed in tunable surface acoustic wave devices\cite{Zdru2022} \textbf{} or magnetoelectric high-overtone bulk acoustic resonators.\cite{Gokhale2023} The main advantage of magnetoelectric composites is the much larger freedom in material selection with the possibility of optimize individually the properties of the piezoelectric and magnetostrictive materials, in addition to the design of device structures that maximize strain coupling. For many practical electronic applications, nanoscale thin film devices are desired. 

\begin{figure}[h]
	\includegraphics[width=14cm]{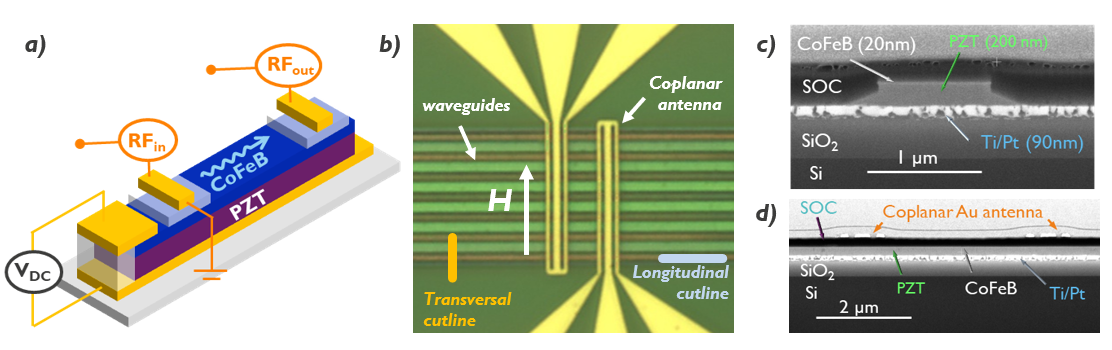}
	\caption{ Electrical scheme of the magnetoelastic devices under test (a) composed by the RF antenna, the applied DC voltage and the 3D representation of a single magnetoelastic waveguide. Optical image of the device showing the parallel waveguides, the coplanar antenna and DC contact (b) with FIB image of the transversal cutline (c) showing the patterned section of the magnetoelastic waveguide and the longitudinal cutline (d), showing the length of the magnetoelastic waveguide and the section of the coplanar antenna.}
	\label{Image1} 
\end{figure}

To measure the magnetoelectric coupling coefficient in such devices, reliable characterization techniques are needed in to assess and optimize the device performance. Typically, the magnetoelectric coupling has been determined for bulk piezoelectric materials,\cite{Pradhan2020, Balinskiy2018} while reports on in nanoscale thin film devices are lacking. Here, we introduce a method to estimate the magnetoelectric coupling coefficient in nanoscale magnetoelastic waveguides by all-electrical propagating spin-wave spectroscopy. Inductive microwave antennas were used to generate and detect propagating spin waves in the waveguides. By applying a DC voltage to electrodes connected to the piezoelectric tin films, mechanical stress is generated in the waveguides and acts on the magnetostrictive layer by changing its internal effective magnetic field. This variation modulates the spin-wave dispersion relation; the modulation can then be used to quantify the internal effective magnetic field and subsequently the coupling strength. A schematic of the devices is depicted in Fig.~\ref{Image1}a.

The device were based on Pb(Zr,Ti)O$_3$ (PZT) and CoFeB thin films as piezoelectric and magnetostrictive layers, respectively. PZT possesses high electromechanical coupling,\cite{Malakooti2013} whereas CoFeB has low magnetic damping,\cite{Bilzer2006} which is beneficial for long-distance spin-wave propagation. Device fabrication started by the deposition of a Ti (20 nm) / Pt (70 nm) bottom electrode on a SiO$_2$ (400 nm)/Si (100) substrate. A 200 nm thick PZT layer was subsequently deposited by pulsed laser deposition (SolMateS BV, SIP-800), followed by a 20 nm thick magnetostrictive CoFeB layer by physical vapor deposition. The piezoelectric/magnetostrictive bilayer was then patterned into 700 nm wide waveguides by e-beam lithography and ion beam etching. The waveguides were planarized by 400 nm thick spin-on carbon (SOC) layer that also provided electrical insulation. Vias in the SOC were opened by plasma etching and filled by Ti (30 nm) /Au (210 nm) to contact the Pt bottom electrode as well as the CoFeB waveguide. Finally, Ti (10 nm) /Au (80 nm) coplanar waveguides and spin-wave antennas were fabricated by a lift-off process. In all experiments below, parallel waveguides separated by a gap of $1.5 \mu m$ were employed to enhance the transmission signal.\cite{Talmelli2019} Fig.~\ref{Image1}b shows an optical micrograph of a final device. 

In the experiments, spin waves were generated and detected by coplanar antennas with ground-signal-ground with individual wire widths of 350 nm - 700 nm - 350 nm. The wires were separated by a 350 nm wide gap. This means that the excitation and detection highest efficiencies were centered around a wavevector of $k \approx 3.1$ rad/$\mu$m, which corresponds to a wavelength of $\lambda \approx 2$ $\mu$m.\cite{Rao2019} The cross-sectional focused ion beam image in Fig.~\ref{Image1}c shows the profile across the magnetoelectric waveguide, while  Fig.~\ref{Image1}d shows a profile along the waveguide and across the ground-signal-ground section of the coplanar antenna.

\begin{figure}[h]
	\includegraphics[width=14cm]{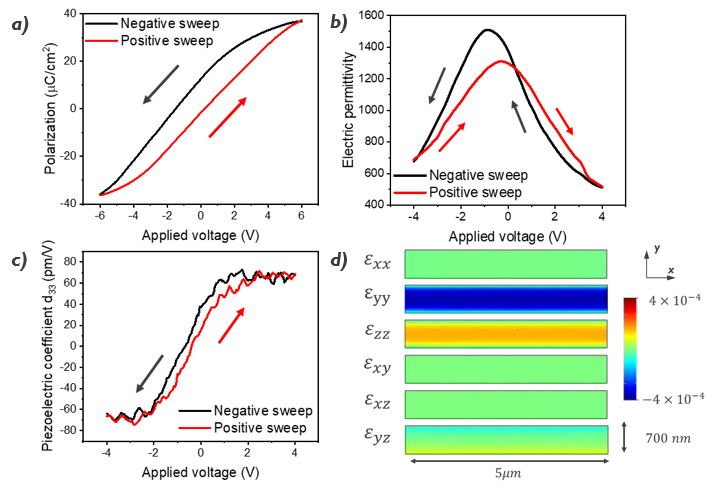}
	\caption{Characterization of the polarization (a), electric permittivity (b) and piezoelectric coefficient (c) of PZT via P-V, C-V and Laser Doppler Vibrometer measurements, respectively. Strain component evaluation in the device simulated by COMSOL multiphysics for an applied voltage of +4V (d) }
	\label{Image2} 
\end{figure}

The polarization of the ferroelectric PZT was characterized by $P(V)$ measurements using separate dot capacitor structures, revealing a remanent polarization of $38$ $\mu$C/cm$^2$ (see ~Fig.\ref{Image2}a). The dielectric permittivity $\varepsilon_r$ was extracted from $C(V)$ measurements, indicating values between 500 and 1500, depending on the applied voltage (see Fig~\ref{Image2}b). The maximum piezoelectric coefficient of the PZT film was measured to be $d_{33}= 66$ pm/V by laser--Doppler vibrometry (see Fig.~\ref{Image2}c), which is the relevant piezoelectric coefficient in the geometry used here.\cite{Ikeda1990}

The mechanical response of the devices, and specifically the strain components due to a DC bias on the CoFeB top electrode, were quantified by finite element simulations implemented in \textsc{Comsol} multiphysics (Fig.\ref{Image2}d), assuming linear piezoelectricity. In this geometry, the DC bias generates an electric field along the $z$-direction between the CoFeB top and Pt bottom electrodes. The simulations indicate that this generates in turn mainly two normal strain components, $\varepsilon_{yy}$ and $\varepsilon_{zz}$, as well as one shear strain component, $\varepsilon_{yz}$, in the CoFeB waveguide. The corresponding magnetoelastic effective field $H_{me}$, is then given by\cite{Kittel_ME_1958,Duflou2017}:

\begin{equation}
	\textbf{H}_{me}=
	-\frac{1}{\mu_0 M_s}
	\begin{pmatrix}
		0\\
		2B\varepsilon_{yy}m_y + B\varepsilon_{yz}m_z\\\
		2B\varepsilon_{zz}m_z + B\varepsilon_{yz}m_y
	\end{pmatrix}\, ,
	\label{Fmel_matrix}
\end{equation}

\noindent with $B$ the magnetostrictive coupling coefficient, $M_s$ the saturation magnetization, and $m_{x,y,z} = M(x,y,z)/M_s$ the normalized magnetization components. During the experiment, an external magnetic bias field was applied in an electromagnet to orient the magnetization of the waveguide. The CoFeB thin films and waveguides show a strong in-plane shape anisotropy and therefore large applied bias magnetic fields comparable to the saturation magnetization would be needed to orient the magnetization along the $z$-direction out of plane. Thus, in-plane configurations have been employed in our experiments. Figure~\ref{Image2}d and Eq.~\eqref{Fmel_matrix} indicate that a the magnetoelastic effective field is maximized when the magnetization is aligned in the $y$-direction. This allows one to access the $\varepsilon_{yy}$ component of the strain, which is the strongest (cf.~Fig.~\ref{Image2}d). For spin waves, this corresponds to the Damon-Eshbach configuration \cite{Eshbach1960, Kalinikos1986, Bhaskar2020}. The magnetoelastic field is then mainly aligned along the $y$-direction with a magnitude of $\mathbf{H}_{me} = -2B\varepsilon_{yy}/\mu_0 M_s \times \mathbf{\hat{y}}$. The saturation magnetization of the film was determined by ferromagnetic resonance measurements to be $M_s=1157$  kA/m, whereas the normal piezoelectric strain of $\varepsilon_{yy}=3.32 \times 10^{-4}$ was obtained from the finite element simulations (\emph{cf.} Fig.~\ref{Image2}d) with +4V of applied bias voltage. Considering a $B =-8.8$ MJ/m$^3$ \cite{Vanderveken2021}, a magnetoelastic field $H_{ME}=5.05$ mT was estimated, with a corresponding magnetoelectric coupling $\alpha_{ME} = 1.26$  mT/V.

\begin{figure}[h]
	\includegraphics[width=16cm]{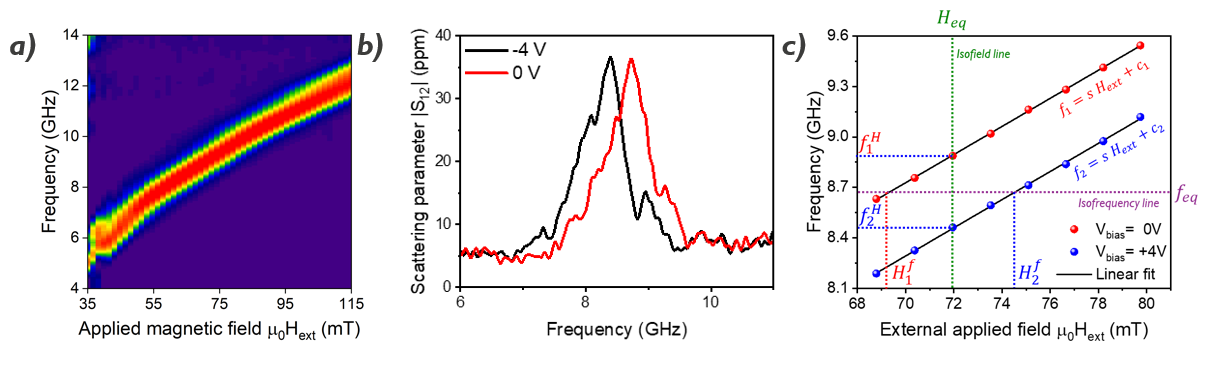}
	\caption{Frequency VS Field map of the transmitted power recorded by propagating spin waves spectroscopy (a). Modulus of the transmission scattering parameter $|S_{12}|$ recorded as a function of the excitation frequency for different applied voltages with a constant applied external field of $\mu_0 H_{ext}=70 mT$ (b). Frequency position of the transmission scattering parameter's peak as a function of the applied magnetic field for different applied voltages (c). }
	\label{Image3} 
\end{figure}

Figure~\ref{Image3}a shows the transmitted power $|S_{12}|$ as a function of frequency and applied magnetic bias field at zero bias voltage. $|S_{12}|$ was recorded by a Keysight E8363B PNA Network Analyzer. In keeping with the spin-wave dispersion relation \cite{Kalinikos1986} an increase in the applied external magnetic field led to a corresponding shift of the resonance frequency towards higher values. Applying a DC bias voltage of $+4$ V resulted in a shift of the the resonance frequency by about 300 MHz with respect to the resonance at zero bias voltage (Fig.~\ref{Image2}b) at a constant applied magnetic field of $\mu_0 H_{ext}=70$ mT.

The frequency of the resonant peak was then extracted as a function of the external magnetic bias field for 0 V and $+4$ V bias voltage (Fig.~\ref{Image3}c). In the studied range between $69$ and $80$ mT, the frequency increased linearly with the external bias magnetic field $H_{ext}$, following $f_{1,2}=s H_{ext}+c_{1,2}$. The subscripts 1 and 2 correspond to 0 V (unbiased) and $+4$ V cases, respectively. The slope $s$ was $83.7 \pm 0.6 \,$ MHz/mT for both bias voltages. The constants $c_{1,2}$ were different, leading to bias-voltage-dependent isofrequency and isofield lines, as shown in Fig.~\ref{Image3}c. Micromagnetic simulations using Mumax3 (Ref.\cite{mumax2014}) were performed for different applied magnetic bias fields in the range shown in Fig.\ref{Image3}c, which indicate that the magnetization distribution does not change significantly with magnetic bias field. This indicates that for identical resonance frequencies $f_{1}$ and $f_2$, the effective magnetic field $H_{eff}$ must be also identical in the waveguide. The total effective magnetic field in the waveguide can be written as $H_{eff}=H_{ext}+H_d+H_{me}(V)$, with $H_d$ the demagnetizing field and $H_{me}(V)$ the magnetoelastic field induced by the applied bias voltage. For an isofrequency line, the condition of identical effective fields leads to the relation
 
\begin{equation}
	H_2^f-H_1^f=H_{me}(V_1)-H_{me}(V_2)=-\Delta H_{me}
	\label{DHm}
\end{equation}

Equation~\eqref{DHm} indicates that the difference in the external magnetic bias field that leads to the identical resonance frequencies for different applied bias voltages is the same as the (change of) magnetoelastic field induced by the bias voltage. Under these specific conditions, a bias voltage of $+4$ V induces thus a magnetoelastic field change of $\Delta H_{me}=5.2$ mT, leading to a (average) magnetoelectric coupling coefficient on the order of $\alpha = \Delta H_{me}/\Delta V = 1.3$ mT/V, which is in great agreement with the expectations provided by the analytical calculation (\emph{cf.} Fig.~\ref{Image2}d).

By combining Eq.~ \eqref{DHm} with the linear expression of the frequency $f=s\, H_{ext}+c$, we can further see that the change in the magnetoelastic field for two applied voltages is directly linked to the different intercepts of the curves. Hence, the magnetoelastic field can be described by

\begin{equation}
	\Delta H_{me}=\frac{c_2-c_1}{s}
	\label{DHm_A}
\end{equation}

These results can be used to correlate the change of the resonance frequency to the change of the magnetoelastic field. At a fixed external applied field $H_{ext}$, we can identify two different frequencies depending on the applied bias voltage that differ by $f_2^H-f_1^H=c_2-c_1$. In combination with Eq.~\eqref{DHm_A}, we obtain

\begin{equation}
	\Delta H_{me}=\frac{f_2^H-f_1^H}{s}
	\label{Df}
\end{equation}

\begin{figure}[h]
	\includegraphics[width=16cm]{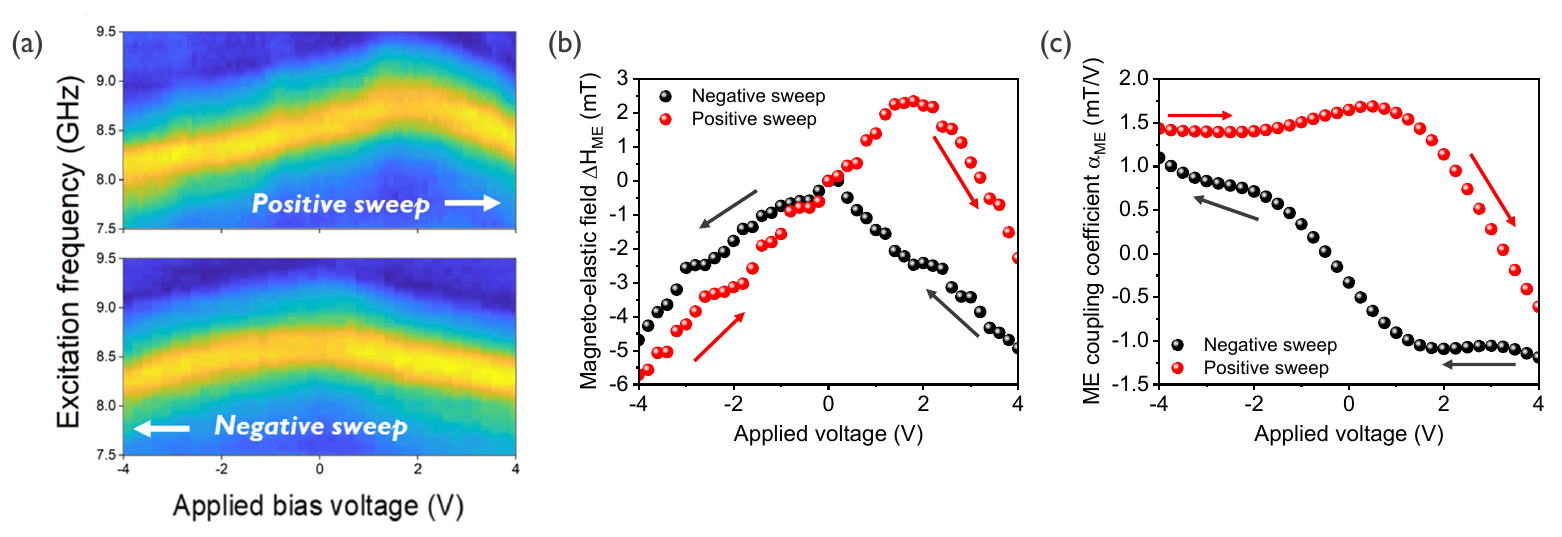}
	\caption{Resonance frequency as a function of the external applied magnetic field for different applied voltage (a). Magnetoelastic field estimation as a function of the applied voltage for the two sweep polarity and for an external applied magnetic field of 70 mT (b). Magnetoelectric coupling coefficient estimation as a function of the applied voltage (c).}
	\label{Image4} 
\end{figure}

This approach can be used to characterize the magnetoelastic field at device level as a function of applied bias. To this aim, the modulus of the transmitted power was recorded at a constant applied magnetic field $\mu_0 H_{ext}=70$ mT as a function of the applied bias voltage between $-4$ V and $+4$ V (positive sweep), followed by a negative sweep from $+4$ V to $-4$ to take into account the hysteretic behaviour of piezoelectric PZT. Figure~\ref{Image4}a shows the transmitted power $|S_{12}|$ as a function of microwave frequency and applied bias voltage for the two sweep directions. The frequency position of the resonance peak was therefore extracted and by using Eq.~ \eqref{Df}, the magnetoelastic field was evaluated to the respect of the frequency for a zero bias voltage, through the relation $\Delta H_{ME}=H_{ME}(V)-H_{ME}(0)$. The results (Fig. \ref{Image4}b) show an hysteresis behavior of the magnetoelastic field depending on the magnitude of the applied voltage, as well as the polarity of the voltage sweep, with peaks in strength of $-5.71$ (for $-4$ V) mT and $+2.35$ mT (for $+1.8$ V), which correspond to an average magnetoelectric coupling coefficient of $\alpha_{ME}=\Delta H_{ME}/\Delta V=1.37$ mT/V. By using the same approach, it is possible to estimate the magnetoelectric coupling coefficient as a function of the applied bias voltage. The results, depicted in Fig. \ref{Image4}c, shows an hysteresis-like coupling coefficient depending on the voltage strength as well as the sweep direction, with a maximum value $\alpha_{ME}=1.69$ mT/V. The found magnetoelectric coupling coefficient, is in line with others already reported in literature of $2.2$\cite{Usami2023}- $2.66$ \cite{Kim2010} mT/V. It is worth to mention that these studies were conducted on bulk single crystal PMN-PT, whereas in our case a 200 nm thin film of polycristalline PZT was employed, leading to similar results. 

In conclusion, we have characterized the magnetoelectric coupling in nanoscale PZT/CoFeB devices using propagating spin-waves spectroscopy. The magnetoelectric coupling was quantified by measuring the shift in frequency induced by the magnetoelastic field generated in the 700 nm wide PZT/CoFeB double layered waveguides. The study provides analytical expressions for the quantification of the magnetoelastic field without the need of any material parameters and are based on the comparison between frequency VS field maps recorded at different applied magnetic fields. The results show nonlinear magnetoelastic field generation linked to the hysteretic nature of the piezoelectric layer, with a maximum value of $\Delta H_{ME}=-5.71$ mT. Finally, the magnetoelectric coupling coefficient was estimated, finding a maximum value of $\alpha_{ME}=1.69$ mT/V and showing, as well, an hysteretic relation with the voltage applied across the piezoelectric layer.\\
D.N. acknowledges the financial support from the Research Foundation – Flanders (FWO) through grant number 1SB9121N. FC, CA, IB, AA and PB acknowledge the financial support received from the Horizon Europe research and innovation program within the project MandMEMS (grant agreement no. 101070536). DN, XW, JDB, CA and  FC acknowledges the support from imec's industrial affiliate program on Exploratory logic, the help from IMEC's lab team for the samples preparation and Chris Drijbooms from IMEC's MCA department for the FIB imaging. The authors acknowledge Matthijn Dekkers and SolMateS's team for the fruitful discussions and the characterization of the piezoelectric films.

\end{document}